\documentclass[aps,prx,10pt,showpacs,amsmath,showkeys,twocolumn,floatfix,amssymb, preprintnumbers, nofootinbib, superscriptaddress,longbibliography]{revtex4-1}
\usepackage{graphicx}
\usepackage{comment}
\usepackage[usenames]{color}
\usepackage{bm}
\usepackage{ifpdf}
\usepackage{floatrow}
\usepackage{makecell}
\usepackage{caption}

\usepackage[normalem]{ulem}
\usepackage[dvipsnames]{xcolor}
\usepackage[utf8]{inputenc}
\usepackage{hyperref}
\hypersetup{
  pdfnewwindow=true,      
  colorlinks=true,        
  linkcolor=PineGreen,    
  citecolor=PineGreen,    
  filecolor=PineGreen,    
  urlcolor=PineGreen      
}
\newcommand{\be}{\begin{eqnarray}}
\newcommand{\ee}{\end{eqnarray}}

\begin{document}

\title{Tunneling time in coupled-channel systems}

\author{Peng~Guo}
\email{peng.guo@dsu.edu}

\affiliation{College of Arts and Sciences,  Dakota State University, Madison, SD 57042, USA}
\affiliation{Kavli Institute for Theoretical Physics, University of California, Santa Barbara, CA 93106, USA}

\author{Vladimir~Gasparian}
\email{vgasparyan@csub.edu}

\affiliation{Department of Physics and Engineering,  California State University, Bakersfield, CA 93311, USA}

\author{Antonio  P\'erez-Garrido}
\email{antonio.perez@upct.es}
\affiliation{Departamento de F\'isica Aplicada,  Universidad Polit\'ecnica de Cartagena, E-30202 Murcia, Spain}

\author{Esther~J\'odar}
\email{esther.jferrandez@upct.es}
\affiliation{Departamento de F\'isica Aplicada,  Universidad Polit\'ecnica de Cartagena, E-30202 Murcia, Spain}

\date{\today}

\begin{abstract}
In present work, we present a couple-channel formalism for the description of   tunneling time of a quantum particle through a composite compound with multiple energy levels or a complex structure that can be reduced to a quasi-one-dimensional multiple-channel system.
\end{abstract}

\maketitle

\section{Introduction}\label{sec:intro}

Tunneling time in quantum mechanics has been one of the long-standing debates in physics \cite{RevModPhys.61.917,RevModPhys.66.217,CHIAO1997345,timeinQMbook}.  The problem has been approached from many
different points of view, and there exists a huge literature on the tunneling problem of electrons through a
barrier, although tunneling times have continued to be
controversial even until now. The most extensively studied is so called B\"uttiker-Landauer time \cite{PhysRevB.27.6178,PhysRevLett.49.1739}, based on the idea to utilize the Larmor precession frequency of the spin (in the weak magnetic fields) as a clock for such time. In this method, the spin is thought to be polarized initially along the direction of travel of the electron (let us say $x$ direction). The rotation of 
the spin, as it traverses the barrier, is then studied by determining the time evolution of its $z$ 
component along the magnetic field transverse to $x$ (let's denote it by $\tau_2$), and along its $y$ direction (let's denote it by  $\tau_1$). Two times, $\tau_1$ and 
$\tau_2$, are then determined as the inverse expectation values of the $y$ and $z$ components, respectively, 
of the Larmor frequency. 
The concept of complex
time (or two time components) in the theory of the traversal time problem of electrons has been studied in many approaches, such as the Green's function (GF) formalism \cite{PhysRevA.69.022106}, the oscillatory incident amplitude and the time-modulated barrier methods  \cite{Markus1,Markus2} and as well as the Feynman path-integral approach, where the
idea of a complex time arises more naturally \cite{dima} (for more details see Ref.~\cite{martin} and references
therein). It is important to notice that the optical analog of the Larmor clock for classical electromagnetic waves based on Faraday effect lead us
also to a complex time \cite{VG1}. Note that in Ref.~\cite{balc}, the optical tunneling times associated with frustrated total internal reflection of a light beam experimentally was investigated. Using the lateral shifts and angular deviations of transmitted and reflected rays as a physical clock,  both components of complex tunneling time $\tau_1 $ and $\tau_2 $ were measured in Ref.~\cite{balc}. The two characteristic interaction times $\tau_1 $ and $\tau_2 $ for classical electromagnetic waves with an arbitrarily shaped barrier are not independent quantities, but are connected by
Kramers-Kronig relations, which relate the real and imaginary components of a causal magnitude.
$\tau_1 $ is proportional to the integrated density of states
for photons and as well as for electrons \cite{PhysRevB.47.2038,PhysRevB.51.6743}.
As for   $\tau_2 $, its interpretation depends on the experiment itself. For example, in the case of the Faraday rotation experiment, $\tau_2 $ is proportional to the degree of ellipticity. 
 In an experiment with disrupted total internal reflection of a light beam \cite{balc}, $\tau_2$ implies superluminal speeds  being highly dependent on boundary conditions and is not associated with the tunneling process. In the presence-time formalism, the second component of time describes the uncertainty of the $\tau_1$ measurement \cite{oscar}.  For a 1D structure coupled to two perfect leads, $\tau_2$ for electrons   is related to Landauer’s conductance through the transmission coefficient   \cite{landauer}.

  The concept of two components of traversal time in  elastic cases was further developed in Ref.~\cite{PhysRevB.51.6743} by taking into account the size of barriers, 
  \begin{equation}
 \tau_2 + i \tau_1    = \frac{d}{d E} \ln  [t (E) e^{2 i k L}]      +  \frac{ r^{(L)} (E)+ r^{(R)} (E)  }{ 4 E}  e^{2 i k L} ,\label{tau5}
 \end{equation}
where $t(E)$ represents the transmission amplitude and $r^{(L/R)}(E)$ denotes the reflection amplitudes for left and right incident waves,
$k=\sqrt{2m E}$ is the wave vector of incoming particle and $L$ stands for the half length of barriers.  The first term  on the right-hand side of Eq.(\ref{tau5}) mainly contains information about the region of the barrier. Most of the information about the boundary is provided by the reflection amplitudes $r^{(L/R)} (E)$   and is of the order of the wavelength, $\lambda$,
over the length of the system $L$:   $ \lambda/L$.  When the  wave packet is  larger than the system size,  the boundary effect becomes  significant for low energies tunneling  and also for small size systems.  The finite-size of sample
effects must be  taken into account and is adequately incorporated by the second term on the right-hand side  of Eq.(\ref{tau5}). Finite-size effects are very important in mesoscopic systems with real leads,  where multiple transmitting modes exist per current path.

The recent advance in attoclock experiments, e.g. \cite{Kheifets_2020,Ramos2020,PhysRevLett.127.133001,natureYu2022,neg1,atoch1}, has shed some light on the possibility  of clarifying some fundamental issues in the debate of tunneling time, i.e., the time the tunneling electron spends under the classically inaccessible barrier. 
The interval of time is measured between
the peak of the electric field, when the bound atomic electron starts
tunneling, and the instant the photoelectron exits the tunnel. According to Refs.~\cite{neg1,atoch1}, timing of the
tunneling ionization was mapped onto the photoelectron
momentum by application of an intense elliptically polarized
laser pulse. Such a pulse served both to liberate an initially
bound atomic electron and to deflect it in the angular spatial
direction. The deflection was taken as a measure of the
tunneling time.  This deflection, as in any direct experiment on the transition time, can be described by a non-stationary
process and must include two time components $\tau_1$ and $\tau_2$ (see Refs. {\cite{PhysRevB.47.2038,balc} for more details). We remark that it 
is not necessarily obvious that experimental measurements of a transit time in such a 
non-stationary process must agree with the calculation obtained on a stationary process. However, as stated above, 
two components of traversal time must exist in any tunneling experiment.  In what follows we will use the concept of two components of tunneling time for the description of tunneling time of a quantum particle through a composite compound with multiple energy levels or a complex structure that can be reduced to a quasi-one-dimensional multiple-channel system.

In  a recent experiment \cite{PhysRevLett.127.133001},  the hyperfine splitting of  the ground state of  a $^{87}$Rb atom was used to  measure single channel elastic scattering tunneling time. Though rigorously speaking, the inelastic effect caused by transition  of $^{87}$Rb  between two hyperfine splitting states of ground state must be considered properly. Due to the small energy gap between two hyperfine splitting states,  $^{87}$Rb can still be well approximated as an elastic system, so that all the  single channel traversal time formalism still applies. The B\"uttiker tunneling time is well described by the energy derivative of scattering phase shift, $\delta  $, 
\begin{equation}
\tau_1 =  \frac{d \delta  }{d E} .
\end{equation}
However, when the inelastic effect becomes more significant, we will show later on in this work that the tunneling time is no longer given directly by the  energy derivative of scattering phase shift, instead it is related to the phase of transmission amplitude, $\phi  $, by
 \begin{equation}
\tau_1 =  \frac{d \phi  }{d E} ,
\end{equation}
where the phase of transmission amplitude is related to both inelasticity, $\eta  $, and scattering phase shift, $\delta  $,  by
\begin{equation}
\phi  = \tan^{-1} \left[ \frac{ \eta  \sin (2 \delta)}{\eta  \cos (2 \delta) +1 } \right]  \stackrel{\eta \rightarrow 1}{\rightarrow} \delta.
\end{equation}
The inelastic effect is described by inelasticity $\eta \in [0,1]$ where $\eta=1$ and $\eta =0$ stand for elastic and totally inelastic cases, respectively.   The transmission amplitude, including inelastic effect,  can be parameterized by 
\begin{equation}
t(E) = \frac{\sqrt{1+\eta^2 + 2 \eta \cos (2\delta)}}{2} e^{i \phi} \stackrel{\eta \rightarrow 1}{\rightarrow} \cos \delta e^{i \delta} .
\end{equation}

The aim of this work is to present the formal  coupled-channel  formalism of tunneling time that can be used to describe inelastic effects properly. Hopefully the extension of tunneling time formalism into inelastic channels offer more opportunities for the examination  of the concept of tunneling time in   experiments. In addition, the coupled-channel formalism also offer a simple and proper mechanism to generate a effective complex potential in a selected subspace of Hilbert space by Hamiltonian projection, see  Ref.~\cite{MUGA2004357}.  As discussed in Ref.~\cite{PhysRevA.69.022106}, a complex potential may generate some interesting effect that is related to the ultrafast propagation of   a quantum wave in absorbing media or barriers.   In particular, numerical modeling of a wave packet propagating in 
the area with effective absorption potential indicates that the arrival time in some cases becomes independent of the travel distance \cite{MUGA2004357}. The latter means that the Hartman effect persists for inelastic scattering too, that is, when the potential becomes non-Hermitian and the scattering matrix is not unitary (for more details see Refs.~\cite{longhi,Hasan2020}).

The coupled-channel formalism of tunneling time may be implemented and realized in various physical systems. In  the present work, we   provide two specific examples that can be described by the same formalism. 
 For the first example, we consider the tunneling of a quantum particle through a composite compound that may exhibit excitations of its internal structures.
A specific   case may be  the scattering process of a electron ($e$) on a molecule,  an atom, or a quantum dot with multiple energy levels (let's refer the composite compound as $X$). The scattering processes involve both  (1) elastic scattering:  $e + X \leftrightarrow e + X$;  and (2) inelastic scattering:   $e + X \leftrightarrow e + X^*$, where symbols $X$ and $X^*$ are used to represent the ground state and excited states of composite compounds, respectively, see Fig.~\ref{scattplot}.    To describe both elastic and inelastic  scattering processes  properly,   a coupled-channel formalism is required. 
 The set of $e+X$ can be referred to as channel 1, while $e+X^*$ is referred to as channel 2,
the elastic scattering is the transition within the same channel, and inelastic scattering describes the transition between two different channels. 
The single-channel formalism of tunneling time that was developed in  Refs.~\cite{PhysRevB.47.2038,PhysRevB.51.6743,PhysRevA.54.4022,PhysRevA.107.032210,Gasparian_2023}  must be generalized to include the inelastic effects such as excitations of compound.  For the second example, we show that a multiple-channel formalism can be realized in a 2D/3D waveguide by confining propagation of quantum particles along one direction. The confinement of quantum particles yield discrete energy eigensolution along transversal directions of waveguide,  which turn a 2D/3D system into a quasi-one-dimensional multi-channel tunneling problem, see e.g. Fig.~\ref{q1Dplot}.   
Due to
the electron's lateral confinement, the propagating modes are mixing
to non-propagating or evanescent modes. The latter decay with the distance, do not carry
a current and do not contribute to the Landauer conductance
of a large sample. However, these evanescent modes are of
paramount importance in Q1D and 2D disordered systems
because they may strongly influence the scattering matrix
elements in an indirect fashion via coupling to propagating
states due to the presence of impurity potentials and due
to tunneling \cite{uzik}.  
 
 \begin{figure}
\begin{center}
\includegraphics[width=0.5\textwidth]{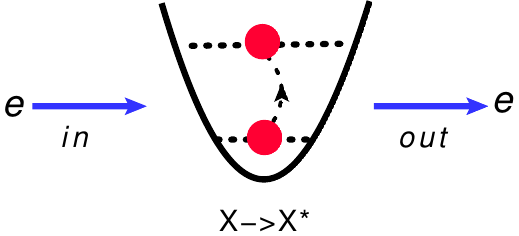}
\caption{ Demo plot of inelastic scattering of  $e + X \leftrightarrow e + X^*$  process. }\label{scattplot}
\end{center}
\end{figure}

We remark that   the  scope of current  discussion in this work is only limited to the short-range interaction   between a quantum particle and a composite compound, the long-range Coulomb potential have not been included yet.  The Coulomb interaction     plays a crucial role in understanding the result of attoclock electron  tunneling ionization time experiment, see e.g. discussion in \cite{doi:10.1080/09500340.2019.1596325,Kheifets_2020}.  For the applications in   tunneling ionization time,    the long-range Coulomb effect need to be incorporated  properly in our future studies.

The paper is organized as follows: In Sec.~\ref{sumFriedelformula}, we present a formal theory of multi-channel scattering, including a generalization of the Friedel formula and tunneling time in multi-channel systems. Sec.~\ref{Example} provides a specific, exactly solvable two-channel system with contact interactions. The physical realization of the coupled-channel formalism in a quasi-one-dimensional waveguide is discussed in Sec.~\ref{secquasi1D}. Finally, discussions and a summary are provided in Sec.~\ref{summary}.

\begin{figure}
\begin{center}
\includegraphics[width=0.8\textwidth]{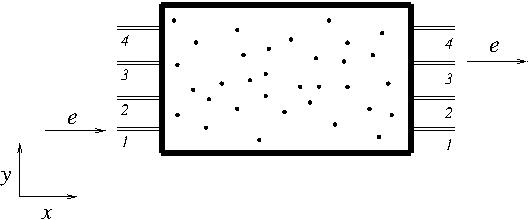}
\caption{ Demo plot of quasi-one-dimensional multi-channel system that is embedded in a 2D waveguide. The black dots represent impurities  placed in the waveguide, the system is confined along $y$-direction.}\label{q1Dplot}
\end{center}
\end{figure}

\section{Formal theory of Friedel formula and tunneling time   in coupled-channel systems}\label{sumFriedelformula}
 In this section,  we present the formal  theory of multi-channel scattering, Friedel formula  and the generalization of tunneling time in coupled-channel systems in a general and formal manner.  Specific examples are given in Sec.~\ref{Example} and Sec.~\ref{secquasi1D}.

\subsection{Formal theory of multi-channel scattering}

The scattering of a non-relativistic  multi-channel system is described by coupled-channel Lippmann-Schwinger equations
\begin{equation}
 | \Psi \rangle =   | \Psi^{(0)} \rangle + \hat{G}^{(0)} (E) \hat{V}   | \Psi \rangle , \label{LSeq}
\end{equation}
 where  
 \begin{equation}
  | \Psi \rangle =  \begin{bmatrix}  | \Psi_1 \rangle  \\  | \Psi_2 \rangle  \\  \cdots    \end{bmatrix}
  \end{equation}
   stands for the   column  vector of  wave functions of multiple-channel  scattering states, similarly $ | \Psi^{(0)} \rangle $ is the column vector of incoming free wave functions of the system. The subscript in $ | \Psi_i \rangle $ is used to label $i$-th particular channel. The multi-channel free Green's function operator is defined by a diagonal matrix
 \begin{equation}
  \hat{G}^{(0)} (E) 
  = \begin{bmatrix} 
  \frac{1}{E- \hat{H}^{(0)}_1} & 0 & \cdots \\
  0 &   \frac{1}{E- \hat{H}^{(0)}_2} & \cdots \\
 \cdots   & \cdots & \cdots
   \end{bmatrix},
 \end{equation}
 where $\hat{H}^{(0)}_i$ is the  free Hamiltonian in the  $i$-th  channel. The interactions between channels is described by $\hat{V} = \{ \hat{V}_{i j} \}$ matrix, where the matrix element $\hat{V}_{i j}$ represents the interaction that couples $i$-th and $j$-th channels. Formally the solution of wave functions is given by
 \begin{equation}
 | \Psi \rangle = \hat{D}^{-1} (E)  | \Psi^{(0)} \rangle   ,  \label{LSeqsolution}
\end{equation}
  with the matrix $\hat{D}(E)$ defined by
 \begin{equation}
  \hat{D}(E) = 1-   \hat{G}^{(0)} (E) \hat{V} .
\end{equation}
 The inverse of $\hat{D} (E)$ is  the matrix of  M{\o}ller operators of a multi-channel system, see e.g. \cite{GOL64}.

 The $S$-matrix of  a multi-channel system is   defined through  the matrix of  M{\o}ller operators  by 
 \begin{equation}
 \hat{S} (E) = \hat{D}(E -i 0 ) \hat{D}^{-1} (E+ i0 ).  \label{Smoller}
 \end{equation}
The  Eq.(\ref{Smoller})  thus yields a relation,
  \begin{equation}
   Im \ln \det [ \hat{D} (E ) ]= - \frac{1}{2 i} \ln \det  [ \hat{S} (E) ].
 \end{equation}
  Assuming that $\hat{D}(E)$ is an analytic function which   only  possesses a physical branch cut lying along positive real axis in complex $E$-plane, we find a dispersive representation of the determinant of $\hat{D}(E)$  in terms of the determinant of $S$-matrix,
  \begin{equation}
    \det [ \hat{D} (E ) ]= N_0  \exp \left( -  \int_0^{\infty} d \lambda \frac{  \frac{1}{2 i} \ln \det  [ \hat{S} (\lambda) ]   }{ \lambda - E } \right), \label{DOmnes}
 \end{equation}
 where $N_0$ is a  constant that cannot be determined by analytic properties of $\hat{D} (E)$ matrix alone. The expression of $\hat{D}(E)$  in Eq.(\ref{DOmnes}) is also known as the Muskhelishvili-Omn\`es (MO) representation \cite{muskhelishvili1941application,Omnes:1958hv}, also see Ref.~\cite{Guo:2022row}.

 The matrix of  scattering amplitude operators, $\hat{T} (E) $, can be defined through coupled-channel Lippmann-Schwinger equations in Eq.(\ref{LSeq}) and Eq.(\ref{LSeqsolution}),
\begin{equation}
 | \Psi \rangle =   | \Psi^{(0)} \rangle - \hat{G}^{(0)} (E) \hat{T} (E)  | \Psi^{(0)} \rangle  ,
\end{equation}
 where 
 \begin{equation}
   \hat{T}  (E)= - \hat{V} \hat{D}^{-1} (E)   .
\end{equation}
 The matrices  of transmission amplitudes and reflection amplitudes   can thus be formally introduced by
  \begin{align}
   & \hat{t}  (E) =1 - \hat{G}^{(0)} (E) \hat{T} (E) =   \hat{D}^{-1} (E) ,  \nonumber \\
   &   \hat{r}  (E) = - \hat{G}^{(0)} (E) \hat{T} (E) .
\end{align}
 The unitarity relation of scattering amplitudes warrants that  $ \hat{t}^\dag  \hat{t}  +  \hat{r}^\dag  \hat{r} =1  $.  Using MO representation of the determinant of  $\hat{D}(E)$ matrix in Eq.(\ref{DOmnes}),  we also obtain a MO representation of the determinant of transmission amplitudes matrix,
   \begin{equation}
    \det [ \hat{t} (E ) ]= N^{-1}_0  \exp\left(   \int_0^{\infty} d \lambda \frac{  \frac{1}{2 i} \ln \det  [ \hat{S} (\lambda) ]   }{ \lambda - E } \right) . \label{tOmnes}
 \end{equation}
The Eq.(\ref{tOmnes})    is   consistent with  results shown in Refs.~\cite{PhysRevA.107.032210,Guo:2022row} in cases of elastic scattering.

\subsection{Friedel formula in multi-channel systems}

 A remarkable relation that connects the integrated Green's function   with the energy derivative of determinant of $S$-matrix is given   in Refs.~\cite{doi:10.1080/00018735400101233,Friedel1958,Faulkner_1977,PhysRevA.6.851.2,Guo:2022row} by 
   \begin{align}
   &  Im   \int_{- \infty}^\infty d x   Tr \left [  \langle x |   \hat{G} (E) - \hat{G}^{(0)} (E) | x \rangle \right ] \nonumber \\
   & = -  \frac{d}{d E}  \frac{1}{2 i } \ln \det  [ \hat{S} (E) ] , \label{Friedelformula}
 \end{align}
 where $\hat{G} (E)$ stands for the full Green's function. The relation in Eq.(\ref{Friedelformula}) is also referred as the Friedel formula. The derivation of Eq.(\ref{Friedelformula}),  in fact, can be made in general, see \cite{PhysRevA.6.851.2,Guo:2022row},  hence the relation in Eq.(\ref{Friedelformula}) is   valid for coupled-channel systems as well. 
 
 In the case of a multi-channel system, $\hat{G} (E)$  now represents   the full Green's function matrix  that satisfies coupled-channel Dyson equations,
 \begin{equation}
  \hat{G} (E) =  \hat{G}^{(0)} (E)  +  \hat{G}^{(0)} (E)  \hat{V}  \hat{G} (E) .
 \end{equation}
 The formal solution of full Green's function matrix is thus given by
   \begin{equation}
  \hat{G} (E) - \hat{G}^{(0)} (E)  =  -   \hat{G}^{(0)} (E)  \hat{T}   \hat{G}^{(0)} (E)  ,  \label{fullGsolution}
 \end{equation}
 and the spectral representation of full Green's function matrix  is
 \begin{equation}
  \hat{G} (E) =  \sum_{\lambda} \frac{  | \Psi (\lambda)  \rangle \langle \Psi (\lambda)  | }{E- \lambda}  .
 \end{equation}

  Assuming that $\hat{G} (E) $ is also an analytic function which   only  possess a physical branch cut lying along positive real axis in complex $E$-plane, we thus obtain
    \begin{align}
  &     \int_{- \infty}^\infty d x   Tr \left [  \langle x |   \hat{G} (E) - \hat{G}^{(0)} (E) | x \rangle \right ] \nonumber \\
      & =    -    \int_0^{\infty} d \lambda   \frac{  \frac{1}{2 i } \ln \det  [ \hat{S} (\lambda) ]  }{(\lambda - E )^2 }  = -  \frac{d}{d E}   \ln \det [\hat{t}(E)] , \label{dGdisp}
 \end{align}
 this resemble the results in cases of elastic scattering  in Refs.~\cite{PhysRevA.107.032210,Guo:2022row}.  We remark that Eq.~(\ref{dGdisp}) not only has been used to define traversal time in quantum tunneling, the Fourier transform of  Eq.~(\ref{dGdisp}) is also related to the second virial expansion coefficient in quantum statistical mechanics, see e.g. Refs.~\cite{Huang_1987,LIU201337}.  In addition, it  also has been found its relevance  in lattice QCD in nuclear/hadron physics recently \cite{Guo:2023ecc,Guo:2024zal}.

\subsection{Traversal time in multi-channel systems} 
For coupled-channel systems,  the definition of  the two components of the traversal time  $\tau_E$   in Refs.~\cite{PhysRevB.47.2038,PhysRevB.51.6743,PhysRevA.54.4022} has to be generalized to
 \begin{equation}
  \tau_E = \tau_2  + i \tau_1  = - \int_{ -   L}^{  L } d x Tr \left [ \langle x | \hat{G} (E) | x \rangle  \right ] , \label{taudef}
 \end{equation}
where $\tau_1$ and $\tau_2$ are  B\"uttiker-Landauer  tunneling time   and the Landauer resistance  respectively.   The $L$ stands for  the half     size of compound. Using Eq.(\ref{dGdisp}) and Eq.(\ref{fullGsolution}), we find
    \begin{align}
   &    \tau_E   =  \frac{d}{d E}   \ln \det [\hat{t}(E)]   -   \int_{- L}^L d x   Tr \left [  \langle x |    \hat{G}^{(0)} (E) | x \rangle \right ]  \nonumber \\
   &  -    (  \int_{L}^\infty + \int_{-\infty}^{-L } ) d x   Tr \left [ \langle x |    \hat{G}^{(0)} (E)  \hat{T}   \hat{G}^{(0)} (E)   | x \rangle  \right ].
 \end{align}

\section{A simple exactly solvable coupled-channel model}\label{Example} 
 
In this section, we consider a simple two-channel model  that represents an electron interacting with  a composite compound $X (X^*)$  through contact interaction potentials. 
 In what follows, we refer to the electron scattering with the ground state of the composite compound $X$ as channel-$1$ and with the excited state of the compound $X^*$ as channel-$2$. The Hamiltonian of the two-channel system is
\begin{equation}
\hat{H} = \begin{bmatrix}  
   \epsilon_1   - \frac{1}{2 \mu_1} \frac{d^2}{d x^2} + V_{1} \delta(x)   & g  \delta(x)  \\  
 g  \delta(x)     & \epsilon_2  - \frac{1}{2 \mu_2} \frac{d^2}{d x^2} + V_{2} \delta(x)    \end{bmatrix}     ,
\end{equation}
 where $\mu_{1,2}$ are the reduced mass of the system in channel-1 and channel-2 respectively, and $\epsilon_{1,2}$ are the threshold factors in channel-1 and channel-2 respectively.  With contact interactions,  negative parity solutions are trivial, hence  our discussion in the following is only restrained to positive parity solutions. $V_{1,2}$ are the strength of contact interaction in channel-1 and channel-2 respectively, and $g$ represents the coupling strength between channel-1 and channel-2.

\subsection{Scattering solutions and parameterization of $S$-matrix}

  The coupled-channel   Lippmann-Schwinger equations with contact interaction potentials are   reduced to a set of algebra equations,
 \begin{align}
 &   \begin{bmatrix}    \psi_1 (x)  \\  \psi_2 (x)   \end{bmatrix}   
  =  \begin{bmatrix}    \psi^{(0)}_1 (x)  \\  \psi^{(0)}_2 (x)   \end{bmatrix}   \nonumber \\
  &+   \begin{bmatrix}    G_1^{(0)}   (x; E)  & 0   \\   0 & G_2^{(0)}   (x; E)   \end{bmatrix}    \begin{bmatrix}    V_{1}  & g   \\   g &  V_{2}  \end{bmatrix}     \begin{bmatrix}    \psi_1 (0)  \\  \psi_2 (0)   \end{bmatrix}  ,
 \end{align}
 where the free-particle Green's function in individual channel is given by
 \begin{equation}
 G_i^{(0)}   (x; E + i 0)  =  \int \frac{d p}{2\pi} \frac{e^{i p x}}{ E -   \epsilon_i - \frac{p^2}{2 \mu_i } - i 0}  = - \frac{i \mu_i}{ k_i} e^{i k_i |x|} .
 \end{equation}
 The relative momentum $k_i$ in  channel-$i$ is related to total energy by
 \begin{equation}
 E = \epsilon_i + \frac{k_i^2}{2\mu_i} .
 \end{equation}

 Two sets of independent solutions are determined by boundary conditions of incoming waves:
 \begin{equation}
(1)  \begin{bmatrix}    \psi^{(0)}_1 (x)  \\  \psi^{(0)}_2 (x)   \end{bmatrix}  =  \begin{bmatrix}   e^{ i  k_1 x}  \\  0  \end{bmatrix} ,  \ \ (2)   \begin{bmatrix}    \psi^{(0)}_1 (x)  \\  \psi^{(0)}_2 (x)   \end{bmatrix}  =    \begin{bmatrix}   0  \\   e^{ i k_2 x }  \end{bmatrix} ,
 \end{equation}
 hence we find
  \begin{align}
(1)  &  \begin{bmatrix}    \psi_1 (x)  \\  \psi_2 (x)   \end{bmatrix}  =  \begin{bmatrix}   e^{ i  k_1 x}  + i \frac{\mu_1}{k_1 } T_{11} e^{i k_1 |x|}   \\   i \frac{\mu_2}{k_2 } T_{21} e^{i k_2 |x|}    \end{bmatrix} ,  \nonumber \\
 (2)  &  \begin{bmatrix}    \psi_1 (x)  \\  \psi_2 (x)   \end{bmatrix}  =    \begin{bmatrix}   i \frac{\mu_1}{k_1 } T_{12} e^{i k_1 |x|}  \\  e^{ i  k_2 x}  +  i \frac{\mu_2}{k_2 } T_{22} e^{i k_2 |x|}    \end{bmatrix} ,
 \end{align}
 where the scattering amplitude $T$-matrix is given by
 \begin{equation}
 T^{-1} (E) =   \begin{bmatrix}    T_{11}  & T_{12}   \\   T_{21} &  T_{22}  \end{bmatrix}^{-1}   =  -     \begin{bmatrix}    V_{1}  & g  \\   g &  V_{2}  \end{bmatrix}^{-1}   -  \begin{bmatrix}  \frac{i \mu_1}{k_1}   & 0   \\   0 &  \frac{i \mu_2}{k_2}\end{bmatrix}  . \label{Tmat}
 \end{equation}
 The two coupled-channel scattering amplitudes can be parameterized by two phase shifts, $\delta_{1,2}$, and one inelasticity, $\eta \in [0,1]$, see e.g. \cite{Guo:2010gx,Guo:2011aa,Guo:2012hv,Guo:2013vsa,Guo:2021uig},
 \begin{align}
T_{11}  &=  \frac{k_1}{\mu_1} \frac{\eta e^{2 i \delta_1} -1}{2i} , \ \ \ \  T_{22}  =  \frac{k_2}{\mu_2} \frac{\eta e^{2 i \delta_2} -1}{2i} ,  \nonumber \\
T_{12}  & =  T_{21} = \sqrt{  \frac{k_1}{\mu_1}  \frac{k_2}{\mu_2} }  \frac{ \sqrt{1- \eta^2 } e^{i  ( \delta_1 + \delta_2)} }{2} .
 \end{align}
 Given explicit expression of scattering amplitudes in Eq.(\ref{Tmat}), the inelasticity and phase shifts can thus be computed by
 \begin{equation}
 \eta = \sqrt{1-  4 \frac{\mu_1}{k_1}  \frac{\mu_2}{k_2} | T_{12} |^2 }, \ \ \ \ \delta_i = \frac{1}{2 i} \ln \left [ \frac{1+ 2 i \frac{\mu_i}{k_i} T_{ii}}{\eta} \right] .
 \end{equation}

 The transmission and reflection amplitudes are defined respectively by
 \begin{align}
& t(E)   =   \begin{bmatrix} 1+ i  \frac{\mu_1}{k_1}   T_{11}  &  i  \frac{\mu_1}{k_1} T_{12}   \\   i  \frac{\mu_2}{k_2}  T_{21} & 1+  i  \frac{\mu_2}{k_2}   T_{22}  \end{bmatrix}   \nonumber \\
&  =      \begin{bmatrix}   \frac{\eta e^{2 i \delta_1} +1}{2}  & i    \sqrt{ \frac{\mu_1 k_2}{ k_1 \mu_2} }  \frac{   \sqrt{1- \eta^2 } e^{i  ( \delta_1 + \delta_2)} }{2}    \\   i  \sqrt{ \frac{\mu_2 k_1}{k_2 \mu_1} }    \frac{   \sqrt{1- \eta^2 } e^{i  ( \delta_1 + \delta_2)} }{2}  &   \frac{\eta e^{2 i \delta_2} +1}{2}  \end{bmatrix}  ,\label{tij}
 \end{align}
 and
  \begin{align}
& r(E)   =   \begin{bmatrix}   i  \frac{\mu_1}{k_1}   T_{11}  &  i  \frac{\mu_1}{k_1} T_{12}   \\   i  \frac{\mu_2}{k_2}  T_{21} &   i  \frac{\mu_2}{k_2}   T_{22}  \end{bmatrix}   \nonumber \\
&  =      \begin{bmatrix}   \frac{\eta e^{2 i \delta_1} -1}{2}  & i    \sqrt{ \frac{\mu_1 k_2}{ k_1 \mu_2} }  \frac{   \sqrt{1- \eta^2 } e^{i  ( \delta_1 + \delta_2)} }{2}    \\   i  \sqrt{ \frac{\mu_2 k_1}{k_2 \mu_1} }  \frac{   \sqrt{1- \eta^2 } e^{i  ( \delta_1 + \delta_2)} }{2}  &   \frac{\eta e^{2 i \delta_2} -1}{2}  \end{bmatrix} \label{rij} . 
 \end{align}
 Using Eq.(\ref{Tmat}), we can verify that the matrix of transmission amplitudes,  $$t(E) = 1 -   \begin{bmatrix}    G_1^{(0)}   (0; E)  & 0   \\   0 & G_2^{(0)}   (0; E)   \end{bmatrix}   T (E),$$ is indeed the inverse of  the matrix of M{\o}ller operators
 \begin{equation}
 t^{-1} (E) = D(E) =1 -    \begin{bmatrix}    G_1^{(0)}   (0; E)  & 0   \\   0 & G_2^{(0)}   (0; E)   \end{bmatrix}   \begin{bmatrix}    V_{1}  & g  \\   g &  V_{2}  \end{bmatrix}.
 \end{equation}
 The determinant of the matrix of transmission amplitudes is given by
 \begin{align}
 \det [t (E)]  & = \frac{\frac{k_1}{\mu_1}\frac{k_2}{\mu_2} }{(\frac{k_1}{\mu_1} + i V_1)  (\frac{k_2}{\mu_2} + i V_2) + g^2 } \nonumber \\
 & =  \frac{\cos (\delta_1 + \delta_2) + \eta \cos (\delta_1 - \delta_2)  }{2} e^{i (\delta_1 + \delta_2)} .
 \end{align}

 The $S$-matrix in parity basis is defined by
  \begin{align}
& S(E)   =   \begin{bmatrix} 1+ 2 i  \frac{\mu_1}{k_1}   T_{11}  &2  i  \frac{\mu_1}{k_1} T_{12}   \\  2 i  \frac{\mu_2}{k_2}  T_{21} & 1+ 2 i  \frac{\mu_2}{k_2}   T_{22}  \end{bmatrix}   \nonumber \\
&  =      \begin{bmatrix}   \eta e^{2 i \delta_1}    & i   \sqrt{ \frac{\mu_1 k_2}{ k_1 \mu_2} }     \sqrt{1- \eta^2 } e^{i  ( \delta_1 + \delta_2)}    \\   i  \sqrt{ \frac{\mu_2 k_1}{k_2 \mu_1} }     \sqrt{1- \eta^2 } e^{i  ( \delta_1 + \delta_2)}   &    \eta e^{2 i \delta_2}    \end{bmatrix}  ,
 \end{align}
 and it satisfies unitarity relation $S^\dag (E) S(E) =1$. We can also show straightforwardly that
 \begin{equation}
 S(E) =t^{-1} (E - i0) t(E+ i 0) ,
 \end{equation}
 where  we have used the relations
 \begin{equation}
  \pm k_i  = \sqrt{2\mu_i (E - \epsilon_i \pm i 0)},
 \end{equation}
 and 
 \begin{equation}
 \delta_{1,2} (E - i 0) = -  \delta_{1,2} (E + i 0) , \ \ \ \   \eta  (E - i 0) =   \eta  (E + i 0) .
 \end{equation}
 The determinant of $S$-matrix is   given by
 \begin{equation}
 \det [S(E) ] = e^{2 i (\delta_1 + \delta_2 )} ,
 \end{equation}
 and hence the MO representation of determinant of transmission amplitudes matrix is 
     \begin{equation}
    \det [ t(E ) ]= N^{-1}_0  \exp\left(   \int_0^{\infty} d \lambda \frac{  \delta_{1} (\lambda ) +  \delta_{2} (\lambda )    }{ \lambda - E } \right) .
 \end{equation}

\subsection{Traversal time in a two-channel system}

 The solution of full Green's functions is determined by coupled-channel Dyson equations that are also reduced to algebra equations for contact interaction,
 \begin{align}
 G(x,x' ;E) & =   G^{(0)}(x-x' ;E)   \nonumber \\
 &  +  G^{(0)}(x ;E)   \begin{bmatrix}    V_{1}  & g  \\   g &  V_{2}  \end{bmatrix}  G(0,x' ;E) ,
 \end{align}
 where 
   \begin{equation}
 G (x, x' ;E)  =  \begin{bmatrix}    G_{11}   (x, x' ; E)  & G_{12}   (x, x' ; E)   \\    G_{12}   (x, x' ; E) & G_{22}   (x, x' ; E) \end{bmatrix}    ,
 \end{equation}
 and
  \begin{equation}
 G^{(0)}(x ;E) =  \begin{bmatrix}    - \frac{ i \mu_1}{k_1} e^{i k_1 |x|}  & 0   \\   0 &  - \frac{ i \mu_2}{k_2} e^{i k_2  |x|}   \end{bmatrix}   .   
 \end{equation}
  Therefore, we find
  \begin{align}
 & G(x,x' ;E)  -   G^{(0)}(x-x' ;E)  \nonumber \\
 &  =  -   G^{(0)}(x ;E)  T (E)   G^{(0)}(x' ;E) .
 \end{align}
 Working out in details,  the trace of integrated Green's function is thus related to the determinant of transmission amplitudes matrix and diagonal terms of reflection amplitudes by
 \begin{align}
&  \int_{-L}^L d x  Tr \left [ G(x,x ;E)  -   G^{(0)}(0 ;E)  \right ] \nonumber \\
 &  = - \frac{d}{d E} \ln \det [t(E)] -  \frac{ \mu_1 r_{11} e^{2 i k_1 L}}{ k_1^2} -  \frac{ \mu_2 r_{22} e^{2 i k_2 L}}{ k_2^2} .
 \end{align}
 Thus the two-component of traversal time is now given by
  \begin{align}
 \tau_E   & = \frac{d}{d E} \ln \det [t(E)]  +  \frac{i \mu_1}{k_1} 2L + \frac{i \mu_2}{k_2} 2L  \nonumber \\
 &    +  \frac{ \mu_1 r_{11} e^{2 i k_1 L}}{ k_1^2} +  \frac{ \mu_2 r_{22} e^{2 i k_2 L}}{ k_2^2} .
 \end{align}
The $\tau_E$ may be interpreted as the total traversal time of a quantum particle through a composite barrier by including all the excitation modes of the composite barrier.

For each individual channel, we can also show that
  \begin{equation}
    \int_{-L}^L d x   G_{ii}(x,x ;E) = \frac{\mu_i}{k_i} \frac{ \partial }{ \partial k_i } \ln \left [  t_{ii} e^{2 i k_i L}   \right ]    +  \frac{ \mu_i r_{ii} e^{2 i k_i L}}{ k_i^2}    , \label{time} 
 \end{equation}
where  the diagonal transmission amplitudes, $t_{ii}$, are 
\begin{equation}
t_{ii}   =\frac{\sqrt{1+ \eta^2 + 2 \eta \cos (2 \delta_i)}  }{2} e^{i  \phi_i  } ,  
\end{equation}
and the phase of $t_{ii}$ is given by  
\begin{equation}
\phi_{i} = \tan^{-1} \left [ \frac{ \eta \sin (2 \delta_ i )}{\eta \cos (2 \delta_ i ) +1 } \right ].
\end{equation}
At the elastic scattering limit: $\eta \rightarrow 1$, the off-diagonal transmission amplitudes approach zero and the diagonal  transmission amplitudes are reduced to elastic expression
\begin{equation}
t_{ii}  \stackrel{\eta \rightarrow 1}{\rightarrow}    \cos  \delta_i    e^{i  \delta_i  }  =  \frac{e^{2 i \delta_i } +1 }{2}  .
\end{equation}
 Consequently, we also find
\begin{equation}
\tau_E = \sum_{i=1}^2 \tau_E^{(ii)},  
\end{equation}
where
\begin{equation}
  \tau_E^{(ii)} =   \int_{-L}^L d x   G_{ii}(x,x ;E), \label{tau11}
\end{equation}
may be interpreted as the two components of traversal time of a quantum particle within $i$-th to $i$-th individual scattering channel in presence of inelastic effect. The B\"uttiker  tunneling time  in $i$-th to $i$-th individual scattering channel is thus explicitly given by
  \begin{equation}
      \tau_1^{(ii)}= \frac{\mu_i}{k_i} \frac{ \partial  \phi_i }{ \partial k_i }    +  \frac{ \mu_i  }{ k_i^2}     \frac{4 k_i L + \eta  \sin  ( 2 \delta_i + 2 k_i L) - \sin ( 2  k_i L ) }{2},\label{real1}  
 \end{equation} 
where the finite-size effect is described by the second term in above expression.

Before concluding this section and for a more complete understanding of the tunneling time in multichannel systems let us introduce so-called  diagonal components of the tunneling time $\tau_E^{(nm)}$. The indices $n$ and $m$ label out-going and incoming scattering channels, respectively, of the system under consideration.
The $\tau_E^{(nm)}$ characterizes the time that a particle spends in both channels between modes $n$ and $m$. This can be defined similarly to the method used above, where the B\"uttiker tunneling time in the $i$-th to $i$-th individual scattering channel was studied
(see Eq. (\ref{real1})). Whether these quantities are by themselves of
physical relevance might well depend on the problem under investigation. While we find that the diagonal elements of $\tau_1^{(nn)}$ 
are positive this is not always the case for the off-diagonal elements $\tau_1^{(nm)}$ (see below).

 For the two-channel case,  the expression for off-diagonal of integrated Green's function   can be written in the form
\begin{align}
\tau_E^{(12)}\equiv\int_{-L}^{L} d x G_{12} (x,x;E)  & =\frac{2 \mu_2}{k_2} \frac{ t_{12}-  r_{12} e^{i (k_1+k_2) L}}{k_1+k_2}, \nonumber \\
\tau_E^{(21)}\equiv\int_{-L}^{L} d x G_{21} (x,x;E) &=    \frac{2 \mu_1}{k_1}    
\frac{t_{21} - r_{21}  e^{i (k_1+k_2) L}}{k_1+k_2}, 
\end{align}
where $t_{nm}$ and $r_{nm}$ are given by Eqs. (\ref{tij}) and (\ref{rij}). Let us now compare, say, the imaginary part of $\tau_E^{(12)}$ (see  Eq.(\ref{taudef}) for the definition of  two components of the traversal time  $\tau_E$) and $\tau_1^{(11)}$ (see Eq.(\ref{real1}))
\begin{align}
 & \rm Im [ \tau_E^{(12)}] = \rm Im [\tau_E^{(21)}] =2\sqrt{\frac{\mu_1\mu_2}{k_1 k_2}}\frac{ \sqrt{1- \eta^2 }} {k_1+k_2} \nonumber \\
 & \times  \sin\bigg(\delta_1 + \delta_2+\frac{(k_1+k_2)L}{2}\bigg)\sin\frac{(k_1+k_2)L}{2}  .
\end{align}
 It is clear that for the selected system parameters the $\rm Im [ \tau_E^{(12)}]$ is negative, in contrast to $\tau_1^{(11)}$, which is always positive.
 Thus one concludes
that, in general, the basic $\rm Im [ \tau_E^{(12)}]$ can not be interpreted as time in the usual sense of the word (see similar discussion in Ref. \cite{PhysRevA.54.4022} about partial density of states and sensitivities in mesoscopic conductors).

\section{Physical implementation  of   quasi-one-dimensional multi-channel systems}\label{secquasi1D}

The physical implementation of  coupled-channel formalism of tunneling time  may be experimentally realized in a quasi-one-dimensional (Q1D) system, which is embedded in a two- or three-dimensional geometry.  A typical example is the propagation of quantum particles in a waveguide in which the electron is confined in the $y$ direction but is
free to propagate in the $x$ direction, see e.g. Fig.~\ref{q1Dplot}. 

In this section, a specific 2D waveguide model is illustrated.
Considering propagation of electron in a  2D waveguide with confinement along $y$-direction and a set of impurities are placed inside the waveguide that play the role of potential barriers. The  system   can thus be described by a simple Hamiltonian,
\begin{equation}
\hat{H} = - \frac{1}{2m} \left(\frac{d^2}{dx^2} +\frac{d^2}{dy^2}\right ) 
+ V_c(y) + V (x, y),
\end{equation}
where  $V_c(y)$ represents the confinement potential along $y$-direction,  the simplest choice of $V_c(y)$ would be infinite square well potential which is zero for
$0 \le y \le L_t$ and infinite elsewhere. The $V(x,y)$ denotes the potential of impurities, which can be modeled by simple contact interactions,
\begin{equation}
V(x,y)=\sum^{N}_{l=1}V_l\delta(x_l-x)\delta(y_l-y), \label{pot}
\end{equation}
  $N$ represents the number of impurities placed inside of the waveguide, and $V_l$ is strength of impurity potential at location of $(x_l, y_l)$.

We remark that the similar 2D mechanism may also be realized in a tight-binding (TB) model with a lattice  size of $L \times L_t$:
\begin{equation}
\hat{H}=\sum^{N}_{i=1}\epsilon_i|r_i \rangle \langle r_i|+g \sum^{N}_{i,j=1}|r_i \rangle \langle r_j|
\end{equation}
where $\epsilon_i$ is the energy of site $i$ and $g$ is the hopping matrix
element. The double sum runs over nearest neighbors.  $(L, L_t)$ are
the length and   the width of the system. The sample is
connected to two semi-infinite, multi-mode leads to the left and
to the right. For simplicity we could take the number of modes in the
left and right leads to be the same $(M)$ and thus the width $L_t$
of this system equals $M$ (for a TB model the number of modes
coincides with the number of sites in the transverse direction).
The analytic solutions of above mentioned   models can be found easily  by
the characteristic determinant  approach, see e.g. Ref.~\cite{Gas:2008}. In spite of the fact that the origins
of these two models are quite different, they are similar in
the sense that their matrix representation for the Hamiltonian
operator has the same structure. Hence they can be
discussed within the framework of the same approach.

The transverse mode wave function $\chi_n(y)$ satisfies a 1D
Schr{\"o}dinger equation:
\begin{equation}
 \left [ - \frac{1}{2m} \frac{d^2}{dy^2} 
+ V_c(y)  \right ] \chi_n(y) = \epsilon_n\chi_n(y),
\end{equation}
being $n$ the sub-band index and $\epsilon_n$ the sub-band
energies.  If the system is confined in transverse direction, such as $V_c(y)$ to be zero for
$0 \le y \le L_t$ and infinite elsewhere, then  solutions in transverse direction are
\begin{equation}
\chi_n(y)=\sqrt{\frac{2}{L_t}}\sin\frac{n\pi y}{L_t}, \ \  \epsilon_n = \frac{\pi^2n^2}{2 m L^2_t}, \ \  n=1,2\hdots.
\end{equation}

Due to confinement along $y$-direction, the electron is only allowed to propagate along $x$-direction  and transits between different $\epsilon_n$  modes. The 2D problem can be reduced to a Q1D scattering problem by integrating over dynamics in $y$-direction. The longitudinal mode of wave function $\varphi_n(x)$,  which is related to total wave function by $\Psi_E(x,y) = \varphi_n(x) \chi_n(y)$, is thus given by a coupled-channel Schr\"odinger equation:
\begin{equation}
\sum_{n'} \left [ -   \delta_{n, n'} \frac{1}{2m} \frac{d^2}{d x^2} + V_{n, n'} (x) \right ] \varphi_{n'}(x) = (E- \epsilon_n )  \varphi_{n}(x).
\end{equation}
  The matrix elements $V_{n,n'}$ are defined by
\begin{equation}
V_{n,n'}  (x) = \int_{0}^{L_t} d y  \chi^*_n(y) V(x, y) \chi_{n'}(y)=\sum^{N}_{l=1}V^{(l)}_{n,n'}\delta(x-x_l),
\end{equation}
with the coupling constant  $V^{(l)}_{n,n'}$ given by
\begin{equation}
V^{(l)}_{n,n'}  ={\frac{2V_l}{L_t}} \sin\frac{n\pi y_{l}}{L_t}\sin\frac{n'\pi y_{l}}{L_t}.
\end{equation}

 The Dyson equation for a Q1D wire can be written in the
form \cite{bag,Gas:2008}
 \begin{align}
& G_{nm}(x,x' ;E)  =   G_{n}^{(0)}(x-x';E)\delta_{n,m}   \nonumber \\
 & +  \sum_{l=1}^N G_{n}^{(0)}(x-x_l;E) \sum^{N}_{q=1} V^{(l)}_{n ,q} G_{q m} (x_l,x';E)  ,
 \end{align}
where $G_{n}^{(0)}(x-x';E)$ is the Green's function in the absence of the defect
potential $V (x, y)$ and obeys the equation
\begin{equation}
\left [ \frac{1}{2m}\frac{d^2}{dx^2} + (E- \epsilon_n)\right ]G^{(0)}_{n}(x-x' ;E) = \delta(x-x') 
\label{eq 1}.
\end{equation}
The explicit form of $G^{(0)}_{n}(x-x' ;E)$ is
\begin{equation}
G^{(0)}_{n}(x-x' ;E)=-\frac{i  m}{k_n}\exp(ik_n|x-x'|)
\label{eq 2}.
\end{equation}
  Here,  $k_n = \sqrt{2m (E-\epsilon_n)}$ is the wave vector.  
The analytic solutions of Dyson equation with contact interaction  $\delta$-potentials  can be found, e.g. by     characteristic determinant approach in  Ref.~\cite{Gas:2009} that is based on the idea of recursively building up the total Green's function. The   transmission and reflection amplitudes of an electron then can be found by using he well-known relations between the scattering amplitudes and GF \cite{Fisher}. Skipping tedious technical details of calculations,  the explicit form   for the matrix elements of Green's function is given by
\begin{align}
G_{nm}(x,x' ;E)  =   G_{n}^{(0)}(x-x';E)\delta_{n,m}   \nonumber \\
 +  r^{(L)}_{nm} \frac{G_{n}^{(0)}(x-x_1) G_{m}^{(0)}(x_1-x')}{\sqrt{G_{n}^{(0)}(0; E)G_{m}^{(0)}(0;E)}},
\end{align}
 $r^{(L)}_{nm}$ is the reflection amplitude for an electron,
incident from the left on the whole system.   Integrating the GF from $[-L, +L ]$ and closely following the procedure presented in Sec.~\ref{Example}, we arrive at a similar expression to Eq.(\ref{time}) for the tunneling time $\tau^{(nn)}_{E}$ in each individual channel. 
 \begin{equation}
\tau^{(nn)}_{E} = \frac{m}{k_n} \frac{ \partial }{ \partial k_n} \ln \left [  t_{nn}  e^{2 i k_n L}  \right ]    +  m\frac{ r^{(L)}_{nn}+r^{(R)}_{nn}}{ 2k^2_n }e^{2 i k_n L}  , \label{time1} 
 \end{equation}
 where $r^{(L/R)}_{nn}$ are for the electrons incident from
the left and right respectively. The left and right reflection amplitudes, $r^{(L/R)}_{nn}$, are not equal to each other in general, $r^{(L)}_{nn} = r^{(R)}_{nn}$ only when the total potential of   system  is symmetric under the spatial inversion. The explicit expressions for reflection and transmission amplitudes are given by Eqs. (14) and (17) in Ref.~\cite {Gas:2009}. In pure 1D system the term related to reflection amplitude/correction term (see Eqs. (\ref{time}) and (\ref{time1})) can be neglected for large systems, for large energies, and in the semiclassical case  (and, of course, if reflection amplitude is negligible). Although there are intriguing general similarities between Eqs. (\ref{time}) and (\ref{time1}), they can be very different in detail. This is mainly due to the fact that in the case of Q1D system (Eq. (\ref{time1})) we are dealing with evanescent modes. The latter in some cases can radically change the physical picture of tunneling time \cite{deo,Gas:2009}.
We remark that the phase factor $e^{2 i k_n L}$ has been absorbed into transmission and reflection amplitudes in  Ref.~\cite {Gas:2009}.

We also remark that   since our formalism is constructed based on analytical properties of $S$-matrix and scattering amplitudes, by analytical continuation, all our results can be formally applied to the evanescent modes as well. The evanescent modes  in Q1D and two dimensional systems may play an important role on the quantum transport properties, such as conductance, average conductance and Hall effect.    The evanescent
modes, like the autoionization states in atomic and molecular systems, appears when the energy $E$ isn’t high enough that in some of the modes vanish in the asymptotic region.   Though the evanescent modes do not carry a current and do not contribute to the Landauer conductance of a large sample \cite{GSuz,emilio,PhysRevB.68.155403,PhysRevB.37.10125,Bagwell_1990,PhysRevB.75.235330,PhysRevB.67.245303},  they may still strongly influence the scattering matrix elements in an indirect fashion via coupling to propagating states due to the presence of impurity potentials and  tunneling \cite{GSuz}.   In some cases they can radically change the physical picture of tunneling time \cite{PhysRevB.67.245303,GSuz}.  Particularly, influence of the evanescent modes on the second term of tunneling time in each channel (Eq. (\ref{time1})) becomes 0 at the Fano resonance due to the evanescent modes (for more details see, e.g.  Refs. \cite{PhysRevB.67.245303,GSuz}). This is in contrast to the truly one dimensional systems, where the correction term is large in quantum regimes.

\section{Summary and outlook}\label{summary}

In summary, we show that the tunneling time of a quantum particle through a composite barrier that displays  excitation of internal structure   and as well as in a quasi-one-dimensional system, which is embedded in a two- or three-dimensional geometry can be described in terms of a coupled-channel formalism. The two components of  the traversal time can be generalized to
   \begin{equation}
 \tau_E    = \frac{d}{d E} \ln \det [t(E)]  + \sum_i   \frac{ 2 i k_i L +         r_{ii} e^{2 i k_i L}}{k_i^2/\mu_i}   ,
 \end{equation}
 where $t(E)$ and $r(E)$ represent the coupled-channel matrices of transmission and reflection scattering amplitudes respectively. The $\tau_E$ in a coupled-channel system may be interpreted as the total traversal time with inclusion of all possible excitation of composite barrier.  The $\tau_E$ can also be related to the traversal time for  the scattering within the same channel,  $\tau_E^{(ii)}$, by
 \begin{equation}
 \tau_E = \sum_i \tau_E^{(ii)},
 \end{equation}
  with
   \begin{equation}
    \tau_E^{(ii)} = \frac{\mu_i}{k_i} \frac{ \partial }{ \partial k_i } \ln \left [  t_{ii} e^{2 i k_i L}   \right ]    +  \frac{ \mu_i r_{ii} e^{2 i k_i L}}{ k_i^2}    .
 \end{equation}
The expression of $\tau_E^{(ii)}$ resembles the single channel expression of traversal time defined in Refs.~\cite{PhysRevB.47.2038,PhysRevB.51.6743,PhysRevA.54.4022} and indicates that the total transverse time is an additive quantity and is obtained by summing over all input modes.  Note that
in spite of the fact that the above equation seems to be self-evident, the analogous theorem has never 
been proved for the multichannel systems.   We also introduce the so-called off-diagonal components of the tunneling time $\tau_E^{(ij)}$. The  off-diagonal components of the tunneling time may characterize the time that particle spends in the both channels between modes $i$
and $j$.
Whether the quantities  $\tau_E^{(ij)}$ are by themselves of
physical relevance might well depend on the problem under investigation. While we find that the diagonal elements of $\tau_E^{(ii)}$ 
are positive this is not always the case for the off-diagonal elements $\tau_E^{(ij)}$.

At last, we remark that   for multi-channel system with more than two channels, the parameterization of scattering amplitudes in terms of phase shifts and inelasticity are not available and may not have a simple form as in two-channel system. Fortunately, the scattering amplitudes can still be modeled in terms of a few dynamical parameters, such potential couplings etc., see e.g. Eq.(\ref{Tmat}). Hence the transmission and reflection amplitudes can be computed in terms of these parameters (see Refs. \cite{Gas:2008,Gas:2009} for more details).

\acknowledgments
P.~G.  acknowledges support from  the College of Arts and Sciences and Faculty Research Initiative Program,  Dakota State University, Madison, SD. V.~G., A.~P.-G. and E.~J. would like to thank UPCT for partial
financial support through “Maria Zambrano ayudas para
la recualificaci\'on del sistema universitario espa\~{n}ol 2021–
2023” financed by Spanish Ministry of Universities with
funds “Next Generation” of EU.  This research was supported by the National Science Foundation under Grant No. NSF PHY-2418937 and in part by the National Science Foundation under Grant No. NSF PHY-1748958.

\appendix

\bibliography{ALL-REF.bib}

\end{document}